\documentstyle[graphicx]{article}
\def\abstract#1{\vskip 7mm 
        \begin{center}{\large Abstract}\par \smallskip
                \begin{minipage}[c]{12cm}
                        \small #1
                \end{minipage}
        \end{center}
}
\def\title#1{\begin{center}{\Large\bf #1}\end{center}}
\def\author#1{\vskip 5mm \begin{center}{#1}\end{center}}
\def\address#1{\begin{center}{\it #1}\end{center}}
\makeatletter
\@ifundefined{lesssim}{}{}
\@ifundefined{gtrsim}{}{}
\def\vereq#1#2{\lower3pt\vbox{\baselineskip1.5pt \lineskip1.5pt
\ialign{$\m@th#1\hfill##\hfil$\crcr#2\crcr\sim\crcr}}}
\makeatother

\begin{document}

\title{%
 Scattering in Two Black Hole Moduli Space
}
\author{%
  Kenji Sakamoto\footnote{E-mail:b1795@sty.cc.yamaguchi-u.ac.jp}
  and \ 
  Kiyoshi Shiraishi\footnote{E-mail:shiraish@sci.yamaguchi-u.ac.jp}
}
\address{%
  Faculty of Science, Yamaguchi University\\
Yoshida, Yamaguchi-shi, Yamaguchi 753-8512, Japan
}
\abstract{
In this work, we discuss the quantum mechanics on the moduli space 
consisting of two maximally charged dilaton black holes. 
We study the quantum effects resulting from 
the different structure of the moduli space geometry in the scattering process. 
}

\section{Introduction}
Recently the study of black hole moduli space has attracted much attention. 
Quantum black holes have been studied by means of quantum fields 
or strings interacting with a single black hole. 
In the past few years the new quantum mechanics of an arbitrary number $N$ of 
supersymmetric black holes has been focused. 
Configurations of $N$ static black holes parametrize a moduli space.
The low-lying quantum states of the system are governed by quantum mechanics 
on the moduli space.
The effective theories of quantum mechanics on a moduli space of
Reissner-Nordstr\"{o}m multi-black holes were constructed 
in \cite{Gibbons}. 
Recently, (super)conformal quantum mechanics are 
constructed on a moduli spaces of five-dimensional 
multi-black holes in the near-horizon limit in \cite{jmas}. 

The motivation behind these works is the hope that information 
of these quantum states will lead to the black hole entropy. 
The quantum states supported in the near-horizon region 
can be interpreted as internal states of black holes, 
and the number of such states is related to the black hole entropy.
The other motivation is the expectation 
that investigation of these moduli spaces will 
lead to the understanding of $\rm{AdS}_2/\rm{CFT}_1$ correspondence 
and unravel some novel features behind the quantum states in multi-black hole mechanics. 

The geometry of black hole moduli spaces was first discussed by Ferrell and 
Eardley in four dimensions \cite{fe}. 
Further the black hole moduli spaces geometry with dilaton coupling 
in $N+1$ dimensions was discussed by Shiraishi \cite{shir}. 
In these works the structure of the moduli space geometry is different 
according to dimensions and values of dilaton coupling. 

In this work, we discuss quantum mechanics on the moduli space consisting of two maximally charged dilaton black holes. 
We study the quantum effect resulting from the different structure of the moduli space geometry in the scattering process. 
In the section 2, we discuss the moduli space structure of two black hole system 
with dilaton coupling in any dimensions. 
In the section 3, we study the quantum mechanics on this moduli space and 
the general view of potential in the (3+1) dimensions. 
In the section 4, we consider the scattering process on the moduli space. 
Then we discuss the quantum effects from the different structures of the moduli space geometry. 
In the section 5, we will give the conclusion and the discussion. 

\section{The Moduli Space Metric for the System Consisting of Maximally Charged 
Dilaton Black Holes}
The Einstein-Maxwell-dilaton system contains a dilaton field $\phi$ coupled to a $U(1)$ gauge field $A_{\mu}$ beside the Einstein-Hilbert gravity.
In the $N+1$ dimensions $(N\ge3)$, the action for the fields with particle sources is 
\begin{eqnarray}
S&=&\int d^{N+1}x \frac{\sqrt{-g}}{16\pi} \left[ R -
\frac{4}{N-1} (\nabla \phi)^{2} - e^{-\{4a/(N-1)\}\phi} F^{2} \right] \nonumber \\
& & \hspace{0.5cm} -\sum^n_{i=1} \int ds_i \left( m_ie^{-\{4a/(N-1)\}\phi} +Q_iA_{\mu}\frac{dx_i^{\mu}}{ds_i}\right),\label{eq:action}
\end{eqnarray}
where $R$ is the scalar curvature and $F_{\mu\nu}=\partial_\mu A_\nu-\partial_\nu A_\mu$. We set the Newton constant $G=1$. 
The dilaton coupling constant $a$ can be assumed to be a positive value.

The metric for the $N$-body system of maximally-charged dilaton black holes has been known as \cite{shir}
\begin{equation}
ds^2=-U^{-2}({\bf{x}}) dt^2+U^{2/(N-2)}({\bf{x}})d{\bf{x}}^2, 
\end{equation}
where
\begin{eqnarray}
U({\bf{x}})&=&(F({\bf{x}}))^{(N-2)/(N-2+a^2)}, \label{eq:U} \\
F({\bf{x}})&=&1+\sum^n_{i=1}\frac{\mu_i}{(N-1)|{\bf{x}}-{\bf{x}}_i|^{N-2}}. \label{eq:F}
\end{eqnarray}

Using these expressions, the vector one form and dilaton configuration are written as
\begin{eqnarray}
A=\sqrt{\frac{N-1}{2(N-2+a^2)}}\left(F({\bf{x}})\right)^{-1}dt, \\
e^{-4a\phi/(N-1)}=(F({\bf{x}}))^{2a^2/(N-2+a^2)}.
\end{eqnarray}
In this solution, the asymptotic value of $\phi$ is fixed to be zero.

The electric charge $Q_i$ of each black hole are associated with the corresponding mass $m_i$ by
\begin{eqnarray}
m_i&=&\frac{A_{N-1}(N-1)}{8\pi(N-2+a^2)}\mu_i,\\
|Q_i|&=&\sqrt{\frac{N-1}{2(N-2+a^2)}}\mu_i, 
\end{eqnarray}
where $A_{N-1}=2\pi^{n/2}/\Gamma(\frac{1}{2}N)$.

The perturbed metric and potential can be written in the form 
\begin{eqnarray}
ds^2=-U^{-2}({\bf{x}})dt^2+2{\bf{N}}d{\bf{x}}dt+U^{2/(N-2)}({\bf{x}})d{\bf{x}}^2, \\
A=\sqrt{\frac{N-1}{2(N-2+a^2)}}(F({\bf{x}}))^{-1}dt+{\bf{A}}d{\bf{x}}, 
\end{eqnarray}
where $U({\bf{x}})$ and $F({\bf{x}})$ are defined by (\ref{eq:U}) and (\ref{eq:F}). 
We have only to solve linearized equations with perturbed sources up to $O(v)$ 
for $N_i$ and $A_i$. 
(Here $v$ represents the velocity of the black hole as a point source.) 
We should note that each source plays the role of a maximally charged dilaton black hole.

Solving the Einstein-Maxwell equations and substituting the solutions, 
the perturbed dilaton field and sources to the action (\ref{eq:action}) with proper boundary terms, 
we get the effective Lagrangian up to $O(v^2)$ for $N$-maximally charged dilaton black hole system
\begin{eqnarray}
L&=&-\sum_{i=1}^n m_i +\sum_{i=1}^n\frac{1}{2}m_i({\bf{v}}_i)^2 \nonumber \\
& & \vspace{1cm} + \frac{(N-1)(N-a^2)}{16\pi(N-2+a^2)^2} 
\int d^N x (F({\bf{x}}))^{2(1-a^2)/(N-2+a^2)}\sum_{i,j}^n
\frac{({\bf{n}}_i\cdot{\bf{n}}_j)|{\bf{v}}_i-{\bf{v}}_j|^2\mu_i\mu_j}
{2|{\bf{r}}_i|^{N-1}|{\bf{r}}_j|^{N-1}} , \label{lag1}
\end{eqnarray}
where ${\bf{r}}_i={\bf{x}}-{\bf{x}}_i$ and ${\bf{n}}_i={\bf{r}}_i/|{\bf{r}}_i|$. $F({\bf{x}})$ is defined by (\ref{eq:F}).
In general, a naive integration in equation (\ref{lag1}) diverges. Therefore, we 
regularize that divergent terms proportional to $\int d^Nx\delta^n(x)/|x|^p(p>0)$ 
which appear when the integrand is expanded must be regularized \cite{InPl}. We set them to zero. The prescription is equivalent to carrying out the following replacement in equation (\ref{lag1})
\begin{eqnarray}
(F(x))^{\frac{2(1-a^2)}{(N-2+a^2)}} \to -1+\left[ 1+
\frac{8\pi(N-2+a^2)}{A_{N-1}(N-2)(N-1)}
\frac{m_a}{|{\bf{r}}_a|^{N-2}} \right]^\frac{2(1-a^2)}{N-2+a^2} \nonumber \\
+\left[1+\frac{8\pi(N-2+a^2)}{A_{N-1}(N-2)(N-1)}
\frac{m_b}{|{\bf{r}}_b|^{N-2}} \right]^\frac{2(1-a^2)}{N-2+a^2}.
\end{eqnarray}
After regularization, the effective Lagrangian for two body system (consisting of black hole labeled with $a$ and $b$) can be rewritten. From this rewriting effective Lagrangian, we obtain the metric of the $N+1$ dimensional moduli space for two-body system as
\begin{equation}
g_{ab}=\gamma(r)\delta_{ab},\label{mmet1}
\end{equation}
with 
\begin{eqnarray}
\gamma(r)=1&-&\frac{M}{\mu}
-\frac{8\pi(N-a^2)}{A_{N-1}(N-2)(N-1)}\frac{M}{r_{ab}^{N-2}} \nonumber \\
&+&\frac{M}{m_a}\left( 1+\frac{8\pi(N-2+a^2)}{A_{N-1}(N-2)(N-1)}
\frac{m_a}{r_{ab}^{N-2}}\right)^{(N-a^2)/(N-2+a^2)} \nonumber \\
&+&\frac{M}{m_b}\left(1+\frac{8\pi(N-2+a^2)}{A_{N-1}(N-2)(N-1)}
\frac{m_b}{r_{ab}^{N-2}}\right)^{(N-a^2)/(N-2+a^2)},\label{mmet2}
\end{eqnarray}
where $M=m_a+m_b$, $\mu=m_am_b/M$, and 
$r_{ab}=|{\bf{x}}_a-{\bf{x}}_b|$. 

\section{Quantum mechanics in two-black hole moduli space}
We consider the quantum mechanics on moduli space. 
The quantization of moduli parameters has been discussed in \cite{TrFe}.

\begin{figure}[hbtp]
\centering
{\includegraphics[width=5.1cm]{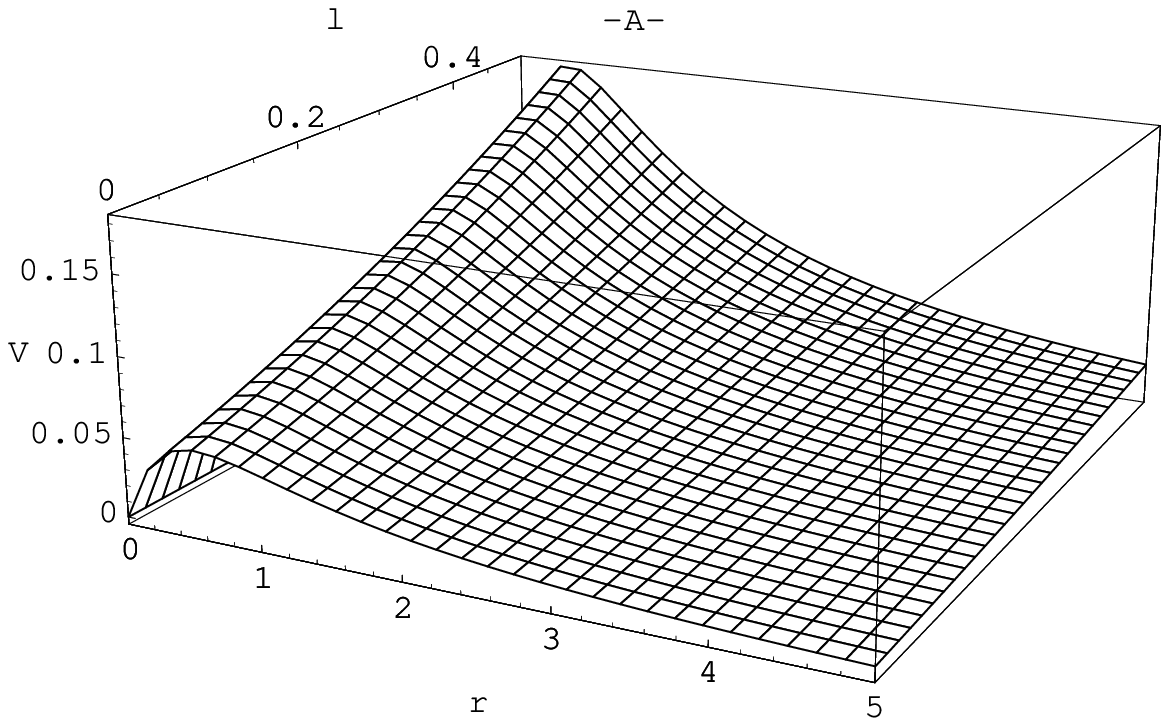}
\hspace{0.05cm}
\includegraphics[width=5.1cm]{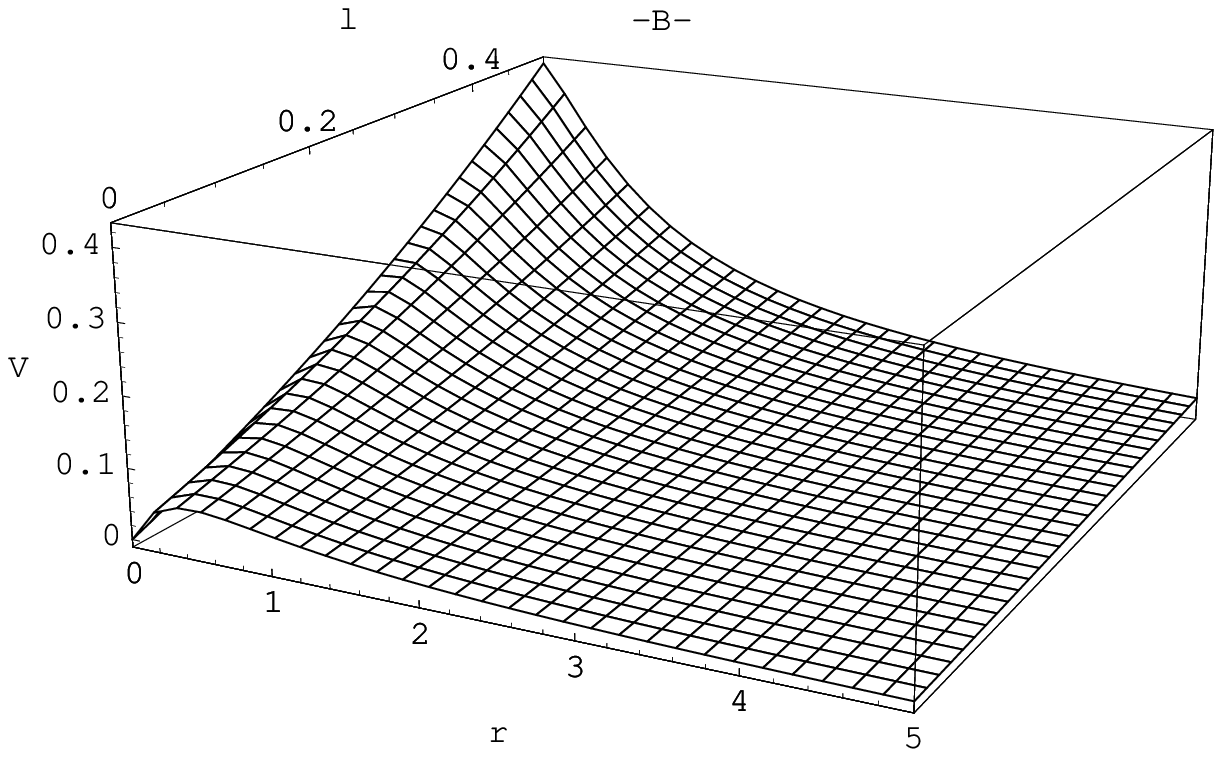}
\hspace{0.05cm}
\includegraphics[width=5.1cm]{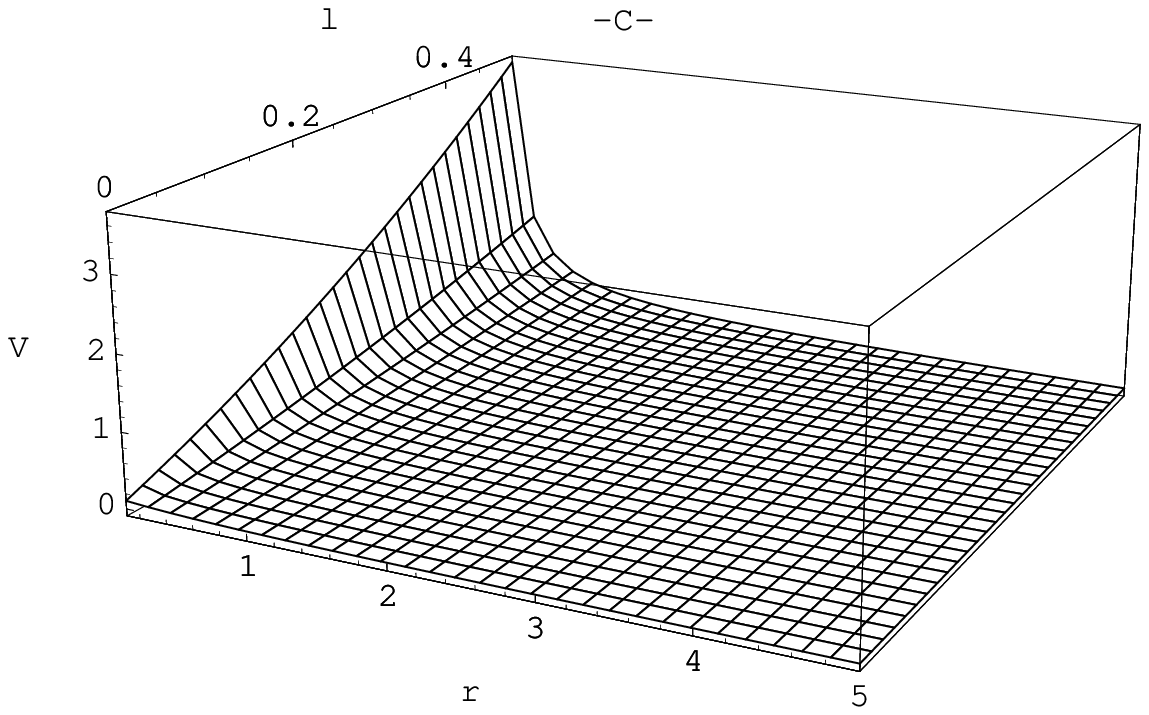}}
\caption{The potential $V$ as the function of $r$ and angular 
momentum $l$, A for $a^2=0$, B for $a^2=1/3$ and C for $a^2=1$.}
\label{ff1}
\end{figure}

Let us introduce a wave function $\Psi$ on the moduli space, which obeys the 
Schr\"{o}dinger equation
\begin{equation}
i\hbar\frac{d\Psi}{dt}=\left(-\frac{\hbar^2}{2\mu}\nabla^2+\hbar^2\xi R_{(MS)}\right)\Psi, \label{scheq}
\end{equation}
where $\nabla^2$ is the covariant Laplacian constructed from the moduli space metric 
and $R_{(MS)}$ is the scalar curvature of the moduli space. 
We assume $\xi=0$ in this paper though this term may be present in most 
general case. \\
To simplify, we fix the case of the (3+1) dimensional case. The partial wave in a stationary state is 
\begin{equation}
\Psi=\psi_{ql}(r)Y_{lm}(\theta,\phi)\exp(-iEt/\hbar),
\end{equation}
where $Y_{lm}(\theta,\phi)$ is the spherical harmonic function 
and $E=\hbar^2 q^2/(2\mu)$.
We redefine the variables as 
\begin{eqnarray}
R&=&\int \sqrt{\gamma} dr, \\
\psi&=&\frac{\chi}{r\sqrt{\gamma}}.
\end{eqnarray}
The Schr\"{o}dinger equation (\ref{scheq}) is rewritten as 
\begin{equation}    
\frac{d^2\chi}{dR^2}+(q^2-V)\chi=0,
\end{equation}
where the potential $V$ is
\begin{equation}
V=\frac{\gamma(r\gamma')'-r(\gamma')^2}{2r\gamma^3}+\frac{l(l+1)}{r^2\gamma}.
\end{equation}
Here $'$ stands for $\frac{d}{dr}$. \\
To study the scattering process, we find the general view of potential on the moduli space. The potential for dilaton coupling $a^2=0$, $1/3$ and $1$ are plotted in Fig.\ref{ff1}.

In the case of $a^2=0$ for all values of $l$ and $a^2=1/3$ for $l=0$, the potential has a maximal value. 
Then the incoming particle into two black hole moduli space are scattered or 
coalesced in the scattering process. 
On the other hand, in the case of $a^2=1/3$ for all values of $l\neq1$ and $a^2=1$ 
all values of $l$, the incoming particle are always scattered away. 
We notice that the value of $a^2=1/3$ is the critical point of moduli structures \cite{shir}. 
For simplicity, we will study the scattering-away processes in the case of $a^2=1/3$ and $a^2=1$. 

\section{Scattering on two-black hole moduli space}
\begin{figure}[hbtp]
\centering
{\includegraphics[width=7.5cm]{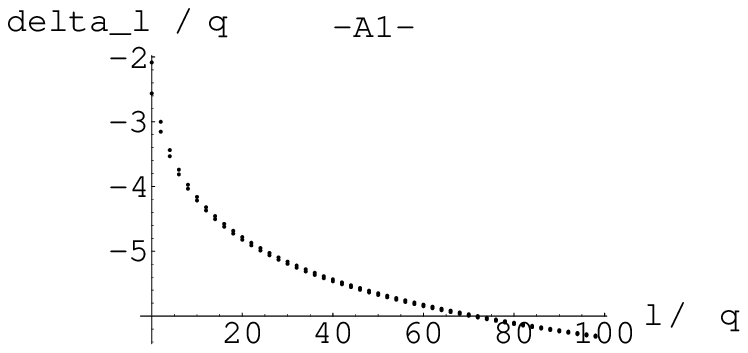}
\hspace{0.05cm}
\includegraphics[width=7.5cm]{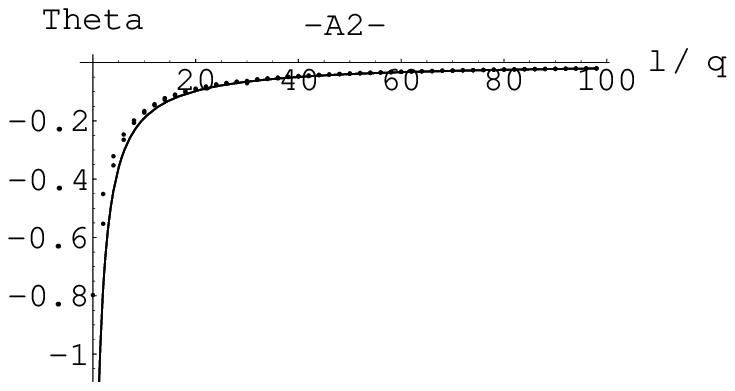}
\hspace{0.05cm}
\includegraphics[width=7.5cm]{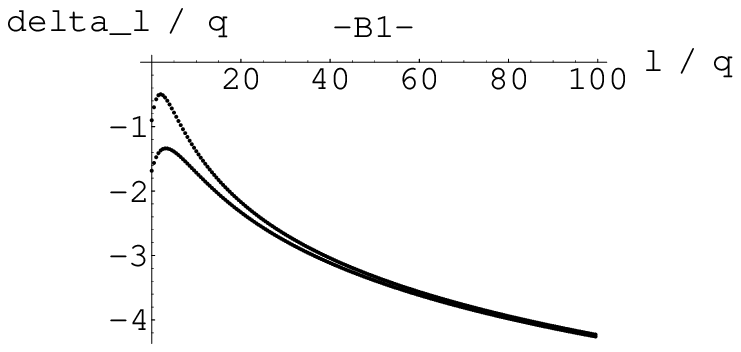}
\hspace{0.05cm}
\includegraphics[width=7.5cm]{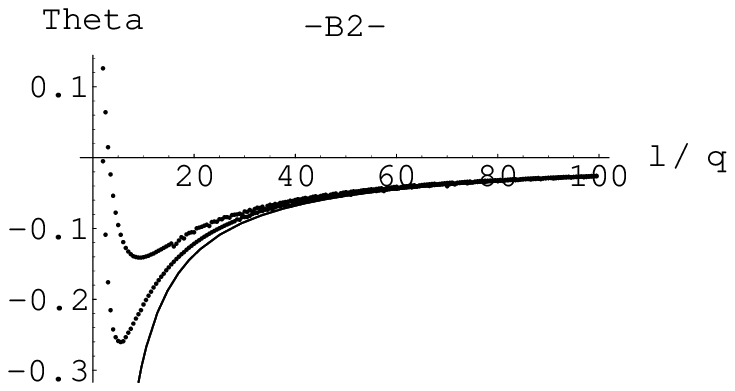}}
\caption{The phase shift and the deflection angle. A1 is a phase shift 
for $a^2=1$, A2 is a deflection angle for $a^2=1$, B1 and
 B2 are phase shift and deflection angle for $a^2=1/3$.}
\label{ff2}
\end{figure}

We consider the case of $a^2=1/3$ and $a^2=1$. 
As the first study of quantum effects, we discuss 
the scattering process in the WKB approximation. 
For the WKB approximation, the phase shift in the scattering process are 
\begin{equation}
\delta_l=-qR_0+\int^{\infty}_{R_0}dR\left(\sqrt{q^2-V(R)-\frac{1}{4R^2}}-q\right)
+\frac{2l+1}{4}\pi,
\end{equation}
where $R_0$ is the solution of $V(R_0)=q^2-\frac{1}{4R_0^2}$. \\
The partial cross section are 
\begin{equation}
\sigma_l=\frac{4\pi}{q^2}(2l+1)\sin^2\delta_l.
\end{equation}
The deflection angle are 
\begin{equation}
\Theta=\pi+\int^{\infty}_{R_0}dR\frac{\partial}{\partial l}\sqrt{q^2-V(R)-\frac{1}{4R^2}}.
\end{equation}
The phase shift $\delta_l$ and the deflection angle
 $\Theta$ for $q=0.1$ and $0.2$ are plotted in Fig.\ref{ff2}. 
The solid line in Fig.\ref{ff2}~A2,~B2 represents the classical deflection angle which are 
obtained from the impact parameter. 
For $a^2=1$, the effects for the different value of $q$ are not found, and 
the deflection angle are almost 
correspond to the classical deflection angle. 
Although, for $a^2=1/3$, we can find that 
the behaviours of the phase shift and the deflection angle depend on 
the incoming particle energy $q$. 
Then the deflection angle are 
different from the classical deflection angle at the small value of $l/q$. 
These effects are interpreted as the quantum effects 
arisen by the difference of the moduli space structures 
in the scattering process. 
Although we used the WKB semi-classical approximation in this analysis, 
the quantum effects can be obtained sufficiently. 

\section{Conclusion}
In this work, we studied the quantum effects on the different moduli structure. 
We obtained the black hole moduli space metrics 
in any dimensions. 
In the (3+1) dimensions, we considered the quantum mechanics on the moduli space, 
and investigated the potential for $a^2=0$, $1/3$ and $1$. 
For the process in which the incoming particle are always scattered away, 
$a^2=1/3$ and $1$, 
we obtained the phase shift and the deflection angle in the WKB approximation. 
Then we revealed the quantum effects in scattering process for $a^2=1/3$. 
We considered that these effects due to the difference 
in the moduli space structure. So, the moduli space structure affect 
to the quantum scattering process. \\
In the further study, we will consider the behavior in the different dimensions and for the other values of dilaton coupling $a^2$. 
The moduli space structures are different in the other dimensions and at the other values of dilaton coupling $a^2$. 
In (4+1) dimensions for $a^2=1$, the moduli space structure are resemble 
the structure in (3+1) dimensions for $a^2=1/3$ \cite{shir}. Then we expect 
that the similar quantum effects are also arisen. 
In this work, we consider the quantum effects in the WKB approximation. 
So, we will more precisely study in quantum mechanically analysis in further study.

\end{document}